\documentclass[final,5p,times,twocolumn]{elsarticle}

\usepackage{lineno,hyperref}
\usepackage{gensymb}

\usepackage{lineno,hyperref}
\usepackage{gensymb}
\usepackage{hhline}
\usepackage{array}
\usepackage{multirow}
\usepackage{booktabs}
\usepackage{tabularx,hhline}
\usepackage{float}
\usepackage{amsmath}
\usepackage{gensymb}
\usepackage{color,soul} 

\modulolinenumbers[5]

\journal{Journal of \LaTeX\ Templates}

\begin{document}

\begin{frontmatter}

\title{Silicon Photomultiplier for Medical Imaging\\ -Analysis of SiPM characteristics-}


\author[NagaiAddress, NagaiSecondaryAddress]{A. Nagai \corref{mycorrespondingauthor}}
\cortext[mycorrespondingauthor]{Corresponding author}
\ead{Andrii.Nagai@unige.ch}

\author[DinuAddress,DinuSecondaryAddress]{N. Dinu-Jaeger}

\author[ParaAddress]{A. Para}

\address[NagaiAddress]{was with the Laboratory of Linear Accelerator \& University Paris 11, CNRS / IN2P3, 91898 Orsay Cedex, France}
\address[NagaiSecondaryAddress]{he is now with Department de physique nucléaire et corpusculaire, Universite de Geneve, 24 Quai E. Ansermet, Switzerland}

\address[DinuAddress]{was with the Laboratory of Linear Accelerator \& University Paris 11, CNRS / IN2P3, 91898 Orsay Cedex, France}
\address[DinuSecondaryAddress]{she is now with ARTEMIS laboratory, CNRS/INSIS, UCA, OCA, F-06304, Nice Cedex, France}

\address[ParaAddress]{is with the Fermi National Accelerator Laboratory, Batavia Il, 60510-5011 USA}

\begin{abstract}
This paper proposes an automatic procedure, based on ROOT data Analysis Framework, for the analysis of 
Silicon Photomultipliers (SiPM) characteristics. 
In particular, it can be used to analyze experimental waveforms, from oscilloscope, containing SiPM pulses acquired 
at different temperatures and bias voltages. 
Important SiPMs characteristics such as: charge distribution, gain, breakdown voltage, pulse shape (rise time and recovery time) and overvoltage can been calculated.
Developed procedure can be easily used to analyze any type of SiPM detectors.
\end{abstract}


\end{frontmatter}


\section{Introduction}
\label{Sec:Introduction}

Nowadays, medical imaging is an increasing field of research because it allows obtaining many morphologic and functional 
information. The main idea of such technology consist's of two parts: in the first step, cancer cells are marked by radioactive isotopes 
and in the second step the emitted particles from isotopes are detected to determine the tumor position. 
Usually, to detect these emitted particles, photon detectors coupled with scintillators are used. 
More details about medical imaging technology are given in Ref.\cite{Dinu}. 

The hand-held intraoperative probes \cite{probes},
used by surgeon in locating and removing tumors, represent one of the interesting applications of medical imaging.
In such case, the arrays of SiPM are today the most promising photon detector candidate because of their 
well adapted characteristics as lightness, compactness, low operating voltage, etc$\dots$ 
However, to obtain a good imaging resolution, all detectors in the array should have uniform electrical characteristics, what it is not a trivial task from the technological point of view.
To study the electrical characteristics of different SiPM devices and their relationship with the technology, the measurement of SiPM as a function of temperature T and bias voltage $V_{bias}$ has to be done, 
because these parameters affect the electrical characteristics of the detectors.
Such kind of measurements result in a huge amount of experimental data. 
Therefore, an automatic procedure for the analysis of SiPM characteristics (e.g. charge distribution, gain, breakdown voltage, pulse shape and overvoltage etc.) 
has been developed and it will be shown hereafter in details. 

\section{The physical principle of SiPM}
The SiPM structure is composed by a parallel array of $\mu cells$ on a common silicon substrate,
where each $\mu cell$ works as a Geiger Mode Avalanche Photodiode (GM-APD) \cite{GMAPD}, connected in series with polysilicon quenching resistor $R_q$.
Each $\mu cell$ works as a digital device, giving a standard output signal independent of number of photons which fired the device (no information on light intensity). 
The SiPM works as an analog device, the output signal being proportional to number of fired $\mu cells$ (it can measure the light intensity).
A detailed description of such devices is give in Ref. \cite{Dinu}.

\section{Experimental}
Experimental measurements have been performed at Silicon Detector facility (SiDet) at Fermilab, USA.

This experimental setup, with fully automatic $N_2$ fulfilling and temperature control, 
gives an opportunity to provide dynamic SiPM measurements in the temperature T range from -175 $^0$C to +55$^0$C. 
Desired T is obtained inside a small box of $\sim 30\times 30 \times 30 cm^2$ called test-cube where the SiPM has been located.
A copper rod, having one extremity in contact with tested SiPM and other one in contact with a vessel filled with $N_2$, serves as a  cold finger used to cool down the samples. 
A platinum resistance thermometer (Pt100) and a resistive heater are mounted close to SiPM detector to control and stabilized the temperature.
To reduce convective heat losses   the test cube has been evacuated to $P \approx 5\times10^{-3}mb$.

The data acquisition system consists of a Miteq amplifier 
(gain = 55dB, bandwidth = 500MHz), Agilent oscilloscope for waveform acquisition 
(a bandwidth of 20MHz has been used to reduce the influence of an electronic noise) 
and Keithley 2400 for SiPM bias supply. All used electronics have been located outside the test-cube, at a room temperature, to 
provide the same experimental conditions (amplifier gain, electronic noise level) independent of T.

An optical fiber conected to a laser is used to illuminate the surface of the tested detector.
The voltage of the laser has been adjusted in such way that only a few photons per pulse were generated. 
More details about experimental setup are given in the Ref. \cite{Dinu}.

Different SiPM devices produced by Hamamatsu HPK have been tested. As an example, the results provided by our analysis Root procedure will be presented for the detector 
MPPC S10931-33-050P-63813, with a total area of $ 3 \times 3 mm^2$ and a $\mu cell$ size of $50 \times 50 \mu m^2$, mounted in a plastic package (produced in 2011 year).

Experimental data have been acquired at different temperatures T ranging from $-175^0C$ up to $+55^0C$ 
in a steps of $10 ^0C$. At each temperature, the device has been operated at twelve biases voltages $V_{bias}$. 
The $V_{bias}$ values at each temperature T have been selected to keep approximately the same overvoltage $( \Delta V = V_{bias} - V_{BD})$ for all temperatures. 
Therefore, the measurements have been performed 
using an overvoltage range from 0.5V to 3.0V with a step of 200 mV. For any operated voltage 5000 waveforms from 
oscilloscope have been saved, each waveform of $5 \mu s$ length.
The waveforms acquisition has been triggered by the electrical signal given by the laser.

\section{Pulse analysing procedure}

To analyze the huge amount of experimental data ($\sim$  70 GB  per  detector) an automatic procedure for data analysis, based on ROOT data Analysis Framework \cite{root}, has been developed. 
This procedure uses experimental waveforms from oscilloscope as input files and creates output Ntuples files with SiPM pulses characteristics for each experimental conditions (T and $V_{bias}$).
The main steps of developed algorithm are: 

\begin{itemize}
\item{\em template creation} to calculate the typical SiPM pulse shape at a given T;
\item{\em pulse finding procedure} to determine time intervals containing SiPM pulses  
(e.g. ``single pulse''\footnote{``single pulse'': a SiPM signal separated by the neighboring pulses by a time interval higher than its recovery time} or 
``train of pulses''\footnote{``train of pulses'': two or more SiPM signals appearing with a time interval between them less than SiPM detector recovery time});
\item{\em template subtraction} to reconstruct all SiPM pulses in the ``train of pulses'';
\item{\em pulse characteristics} to calculate the main SiPM pulse parameters.
\end{itemize}
All these steps will be presented in details in the following.

\begin{figure}
\begin{center}\includegraphics[%
  width=9cm,
  keepaspectratio]{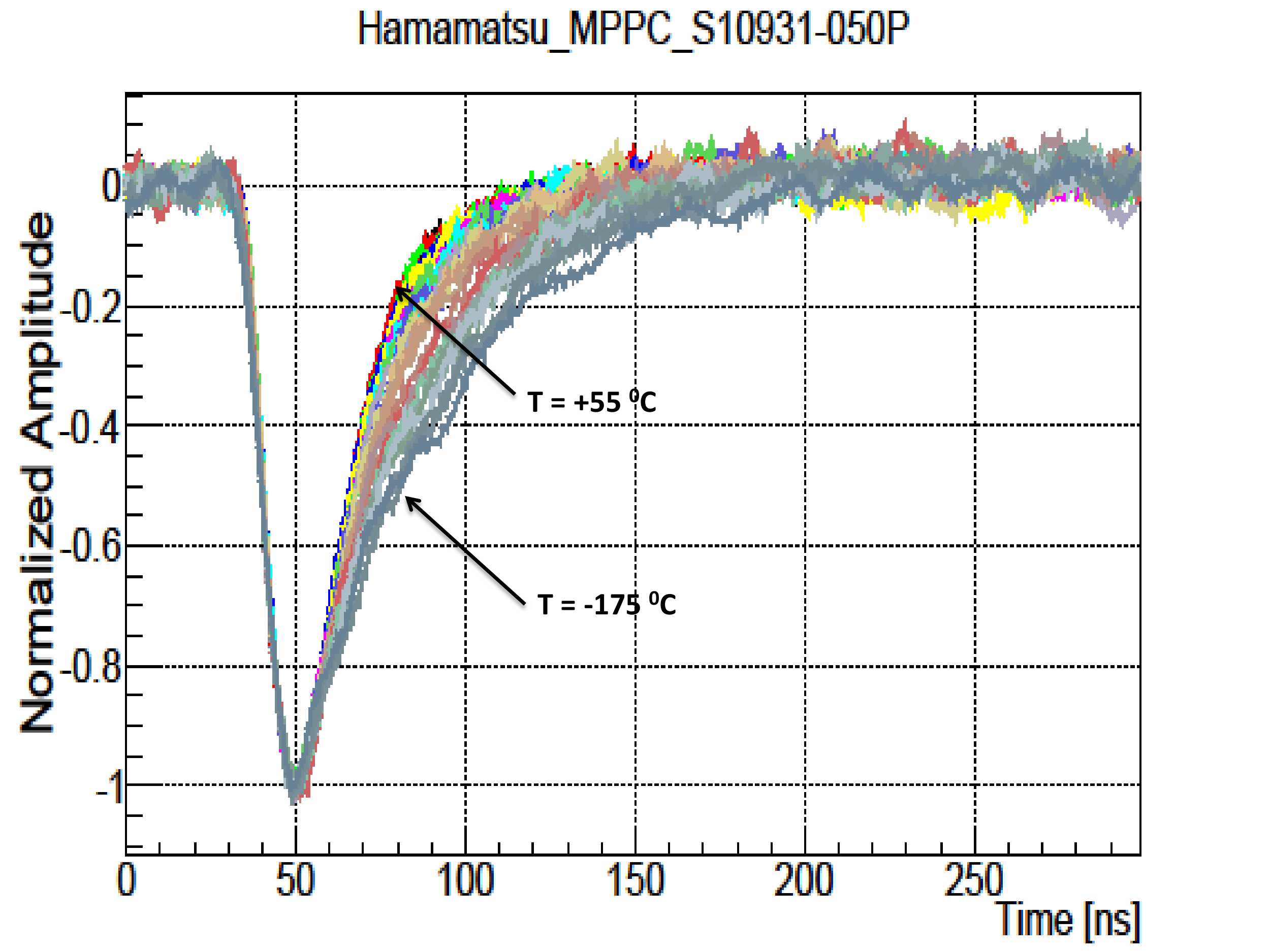}
\end{center}
\caption{Colection of templates at different temperatures}
\label{Template}
\end{figure}

\paragraph{Template creation.}At a given T, the SiPM pulse shape is independent of the applied $V_{bias}$. However, because of T dependence of $R_q$ \cite{Dinu} we expect that the pulse shape changes with T.
Therefore, for each T the typical normalized pulse of SiPM has been determined (what will be call in the following ``template'')
by selecting ten ``single SiPM pulses'' determined by the laser light.  
To remove electronic pedestal, the baselines of each pulse have been calculated as mean amplitude values over an interval of 40ns length before SiPM pulse leading edge.
After baseline subtraction, all selected pulses have been normalized and the template was created as an average over them.
The collection of normalized templates at different T is presented in Fig.\ref{Template}.
Even if the shape of our pulses are slightly modified by our read-out electronic (e.g 20MHz oscilloscope bandwidth), we can distinguish that each
pulse is characterized by a fast leading edge (e.g. rise time) and a slower trailing edge (e.g. recovery time). 
In the limits of our experimental errors, rising edge shows no T dependence, while the recovery time changes from 45ns at $+55^0C$ to 160ns $-175^0C$ 
because of temperature dependence of polysilicon quenching resistor $R_q$.
\paragraph{Pulse finding.}The pulse finding procedure has been used to determine all time intervals which can contain the SiPM pulses. 
This procedure calculates a trigger level as $5\times \sigma$, where $\sigma$ is the standard deviation related to the average performed over all amplitude values of the waveforms acquired at a given T and $V_{bias}$.
In this step of analysis, each found waveform pieces will be considered as a ``train of pulses'' independent if it contains a ``single pulse'', ``train of pulses'' or high frequency noise with amplitude exceeding the trigger level.

\begin{figure}
\begin{center}\includegraphics[%
  width=8.2cm,
  keepaspectratio]{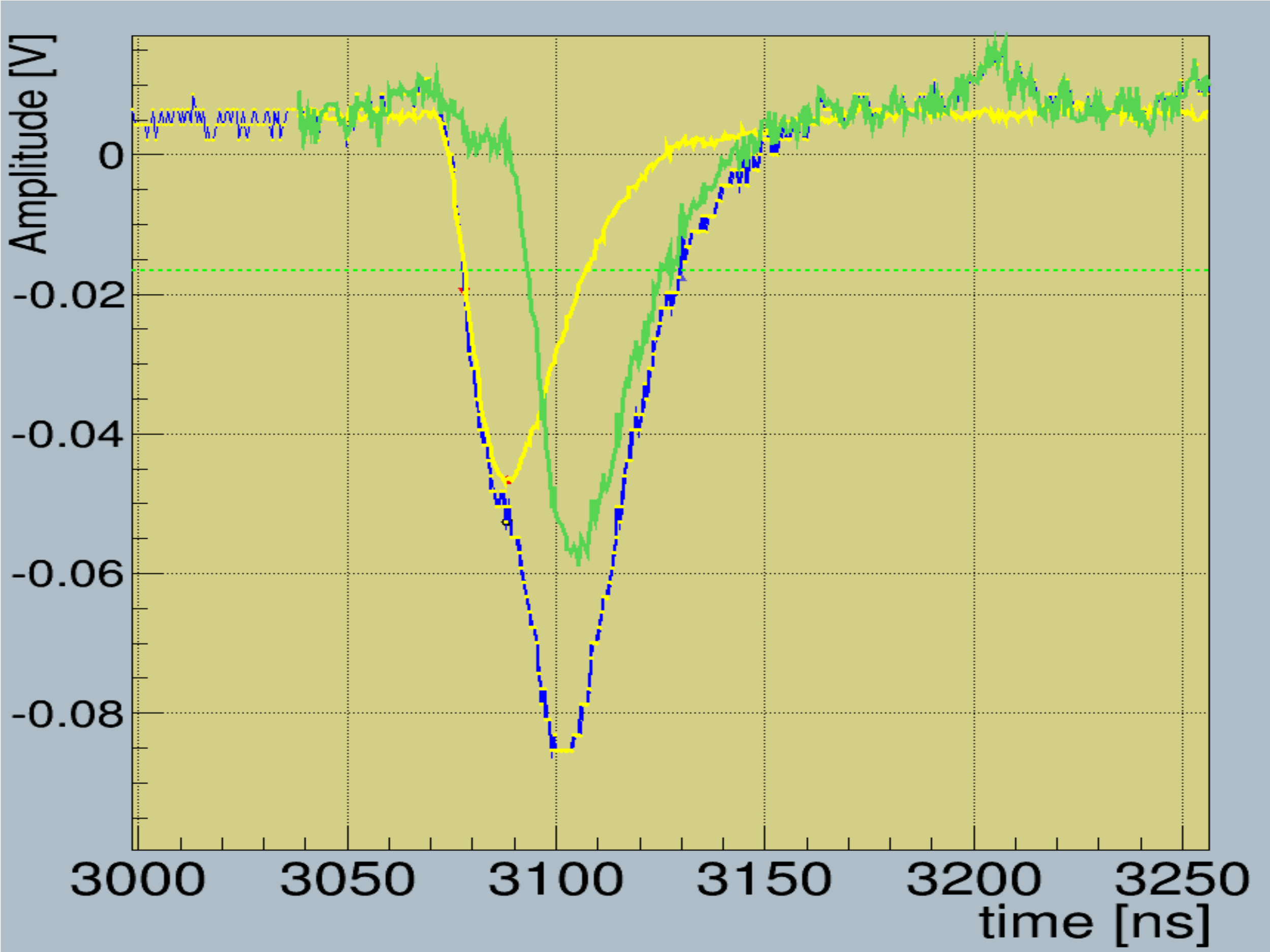}\end{center}
\caption{Example of a part of waveform with a ``train of pulses''(blue line),
template which fits rising edge (yellow line) and waveform after first pulse subtraction (green line)
, at $T = +55^0C$, $V_{bias}$ = 72.52V and $\Delta V$ = 1.2V}
\label{Two_pulses}
\end{figure}

\begin{figure}
\begin{center}\includegraphics[%
  width=8.2cm,
  keepaspectratio]{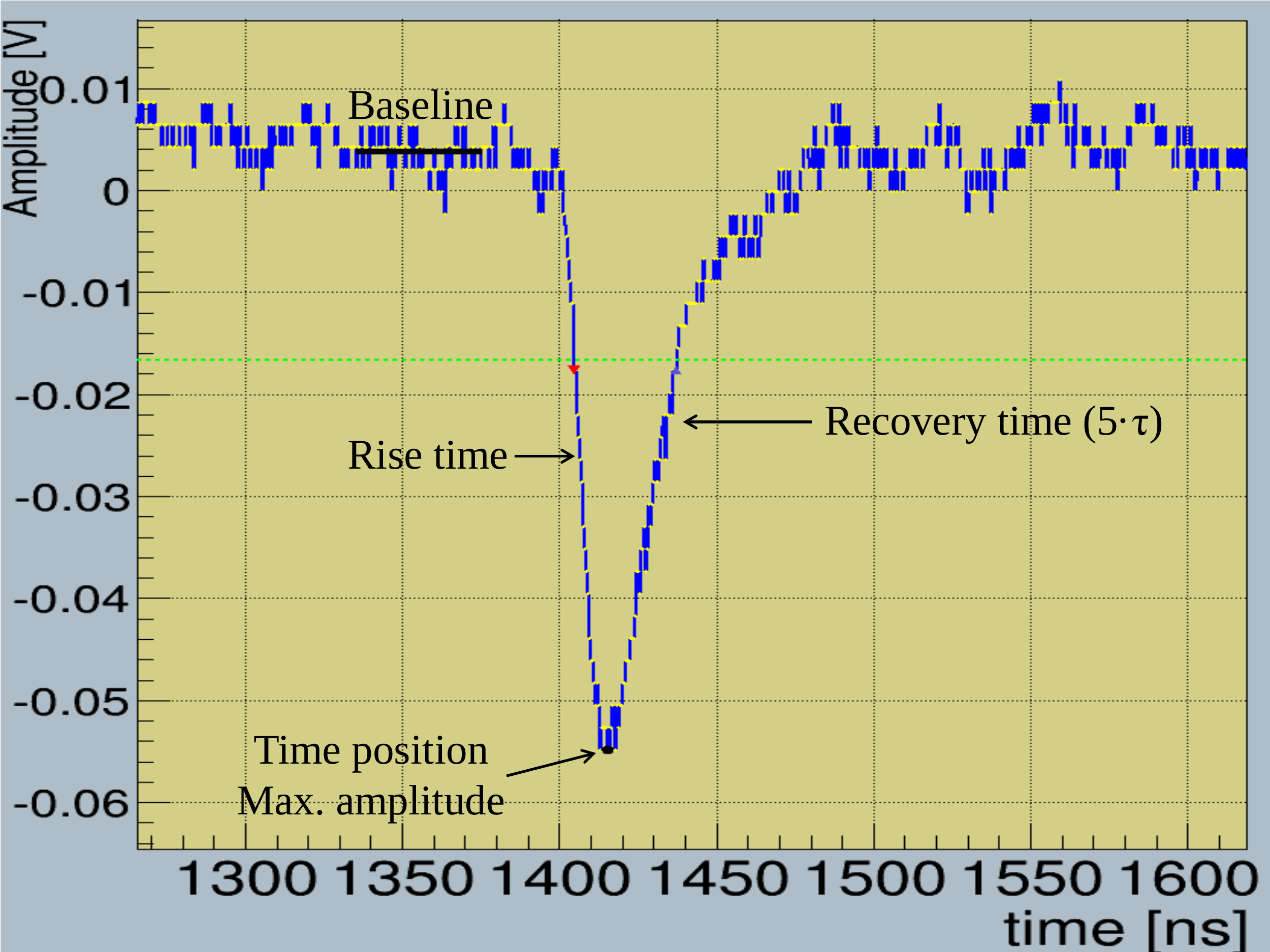}\end{center}
\caption{Typical ``single'' SiPM pulse and its main parameters, at $T = +55^0C$, $V_{bias}$ = 72.52V and $\Delta V$ = 1.2V}
\label{Single_pulse}
\end{figure}

\paragraph{Template subtraction.}To reconstruct SiPM pulses inside of a ``train of pulses'' determined in the previous step, the template subtraction procedure has been developed and used. 
The main idea of this procedure is to fit SiPM pulse rising edge by template rising edge and then to subtract whole template from the fitted pulse.
An example of this procedure is presented in Fig.\ref{Two_pulses}, where blue line represents a part of experimental waveform which contains 
a ``train of two pulses'', yellow line represents the template which fits rising edge 
and green line represents the second pulse which was reconstructed thanks to template subtraction procedure applied to the first pulse in the ``train of pulses''.
This procedure has been developed and used in such way that it can reconstruct all pulses independent of the number of pulses in a ``train of pulses''.
\paragraph{Pulse characteristics.} For each pulse the main parameters such as: {\em time position, $t_{before}$\footnote{time distance between analyzed pulse and the previous one}, $t_{after}$\footnote{time distance between analyzed pulse and the next one}, baseline, amplitude, rise time, recovery time, chi-square $\chi^2$ and charge Q} have been calculated.
A typical ``single'' SiPM pulse and its main parameters are presented in Fig.\ref{Single_pulse}.
The {\em time position} is given by the time of the point with maximum amplitude (in absolute value) of the SiPM signal.
The {\em $t_{before}$} and {\em $t_{after}$} have been calculated as a time interval between analyzed pulse and neighboring pulses. 
The {\em baseline} has been calculated as an average amplitude over a time interval of 40ns length before starting pulse leading edge.
The pulse {\em amplitude} has been calculated as a difference between maximum amplitude value (in absolute value) and calculated baseline.
The {\em rise time} has been calculated as a time during which pulse amplitude increases from $10\%$ up to $90\%$ of a maximum value (in absolute value).
The {\em recovery time}, has been calculate as $5\cdot \tau$, where $\tau$ is a time constant calculated from exponential fit $( f(t,a,\tau) = a\cdot \exp {\left( - t/ \tau \right) }$, where a - is a free parameter) of a pulse falling edge.
The {\em chi-square $\chi^2$}, has been calculated from a comparison between normalized SiPM pulse and template at a given T as:
\begin{equation}
 \chi^2 = \frac{ \sum_{t=0}^{N_{points}} \left( y^{template}(t) - \frac{y^{pulse}(t)}{ \left|A^{pulse} \right| } \right)^2 }{N_{points}} 
\end{equation}
where $y^{template}(t)$ - template amplitude value at time t,  $y^{pulse}(t)$ - pulse amplitude value at time t, $A^{pulse}$ - pulse amplitude, $N_{points}$ - number of experimental points inside of a typical SiPM pulse length at a given T.
The {\em charge Q} for each pulse has been calculated  as:
\begin{equation}
 Q= \int I(t)dt = \frac{1}{R \times G_{amplifier}} \int V(t)dt
\end{equation}
where I(t) - current at time t, $G_{amplifier}$ - amplifier gain ($G_{amplifier}$ = 55dB), R - amplifier input impedance (R = 50$\Omega$ ), V(t) - pulse amplitude value at time t. 
The integral has been calculated in a time range equal to pulse rise time plus pulse recovery time by using the numerical trapezoidal rule \cite{integral}. 

All calculated pulses parameters at a given T and $V_{bias}$ have been saved into output Ntuples root binary files (one file at a given T and $V_{bias}$).
To work with Ntuple files in real time, an user friendly interface was developed, based on a ROOT framework.
Moreover, it allows also to visualize the experimental waveforms and to check the full analysis procedure.
\section{SiPM parameters}
\begin{figure}
\begin{center}\includegraphics[%
  width=7.2cm,
  keepaspectratio]{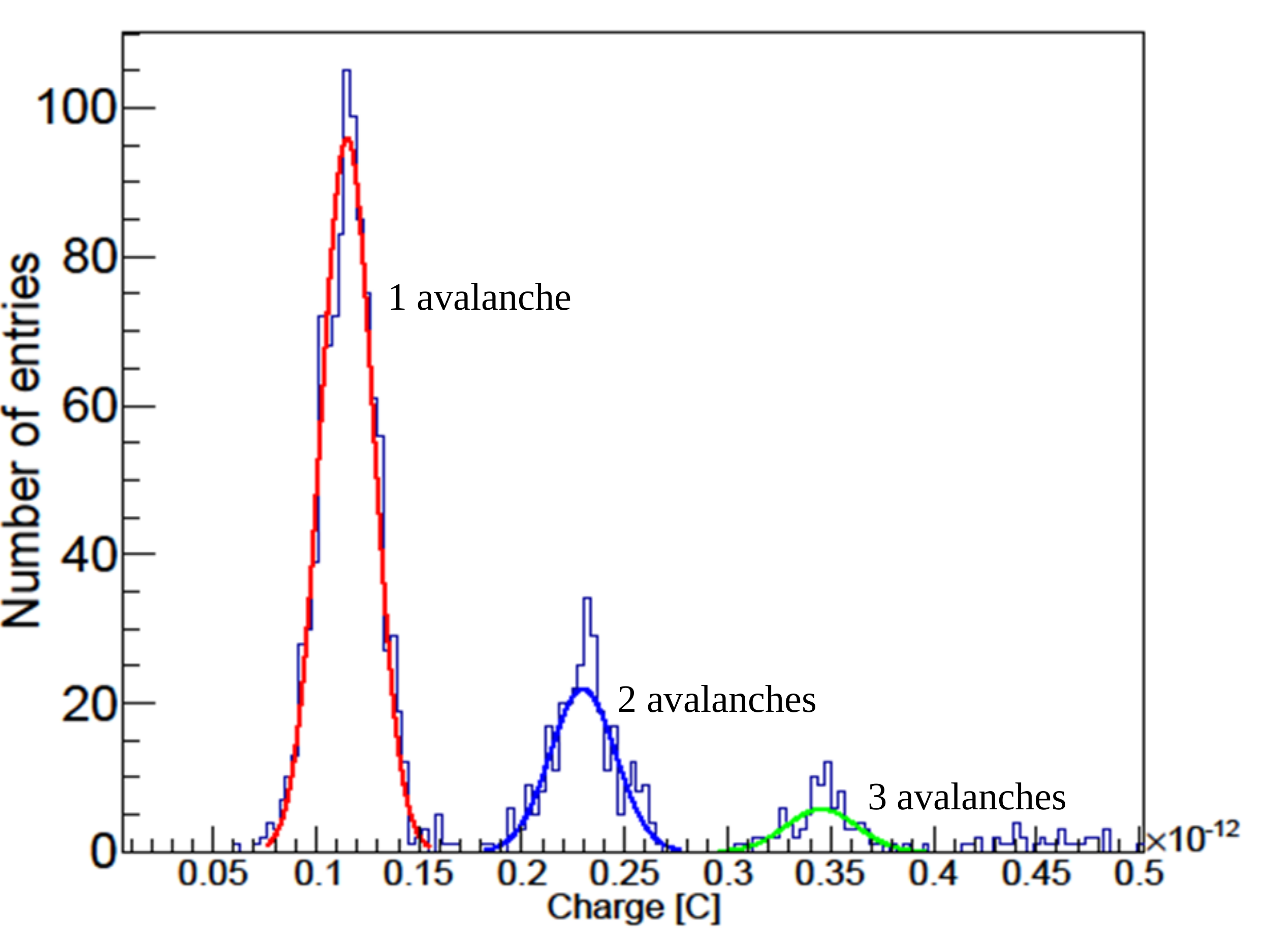}\end{center}
\caption{Typical charge histogram when only ``single'' pulses are selected, at $T = -25^0C$, $V_{bias}$ = 68.4V and $\Delta V$ = 1.2V}
\label{Charge}
\end{figure}

In the first approximation pieces of waveforms which contains only SiPM pulses were used to calculate SiPM detector parameters.  
Such pieces have been determined by using cuts to $\chi^2$, rise time and times to neighboring pulses (e.g. $t_{before} \:\ and \:\ t_{after}$). 
An example of a typical charge distribution corresponding to a given experimental conditions (e.g. $T = -25^0C$, $V_{bias}$ = 68.4V, $\Delta V$ = 1.2V) is presented in Fig.\ref{Charge}.
The plot shows three Gaussians appropriates to signals determined by one, two and three avalanches.  
The charge corresponding to one avalanche signal has been calculated as a mean value of the first Gaussian fit. 
To determine the right fit range a combined first and second derivative method \cite{peaksfinding} was developed and used.

\begin{figure}
\begin{center}\includegraphics[%
  width=7cm,
  keepaspectratio]{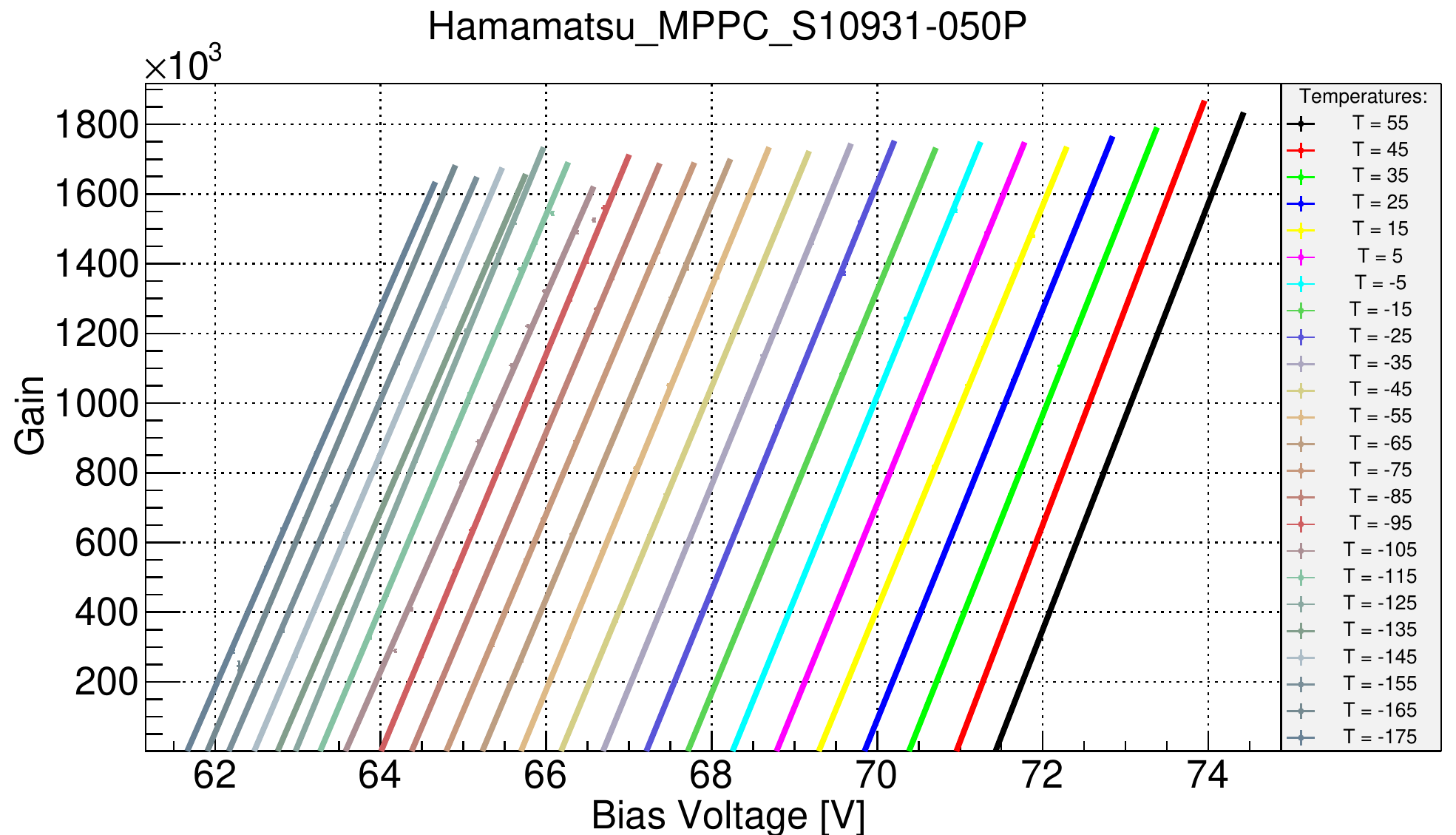}\end{center}
\caption{Gain as a function of $V_{bias}$ in the temperature range from -175$^0$C to +55$^0$C}
\label{Gain3x3}
\end{figure}

The gain G as a function of $V_{bias}$ has been calculated as: $G=Q/e^-$ ($e^-$ - electron charge) and it is presented in Fig.\ref{Gain3x3}. 
At a given T, the G increase linearly with $V_{bias}$ as expected  (e.q. $ G= \left[ C_{\mu cell} \times  \left( V_{bias} - V_{BD} \right) \right] /e^-$, $ C_{\mu cell}$ - $\mu cell$ capacitance).
We can note that, if $V_{bias}$ is kept constant, the G is changing of about  $6 \% /^0C$. 

\begin{figure}
\begin{center}\includegraphics[%
  width=8cm,
  keepaspectratio]{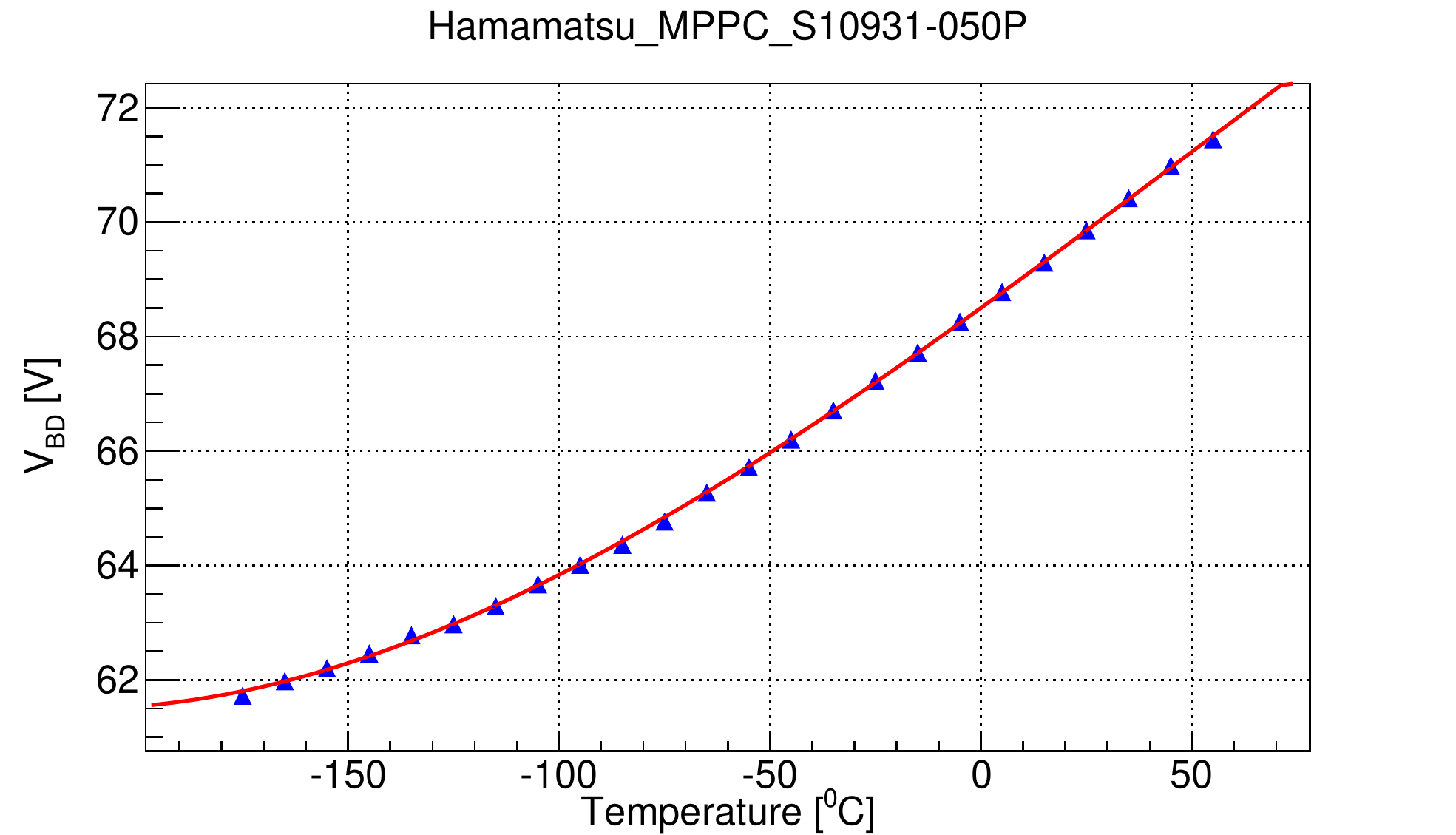}\end{center}
\caption{$V_{BD}$ vs. T}
\label{Vbd_3x3}
\end{figure}

The breakdown voltage $V_{BD}$ at a given T has been determined from the intersection of the linear fits with abscise axis in the G vs $V_{bias}$ plot. 
The $V_{BD}$ as a function of T is presented in Fig.\ref{Vbd_3x3}.
We can observe that $V_{BD}$ decreases linearly with decreasing T up to $-55^0C$. 
Going down in T the $V_{BD}$ present nonlinear temperature dependence.

\begin{figure}
\begin{center}\includegraphics[%
  width=7cm,
  keepaspectratio]{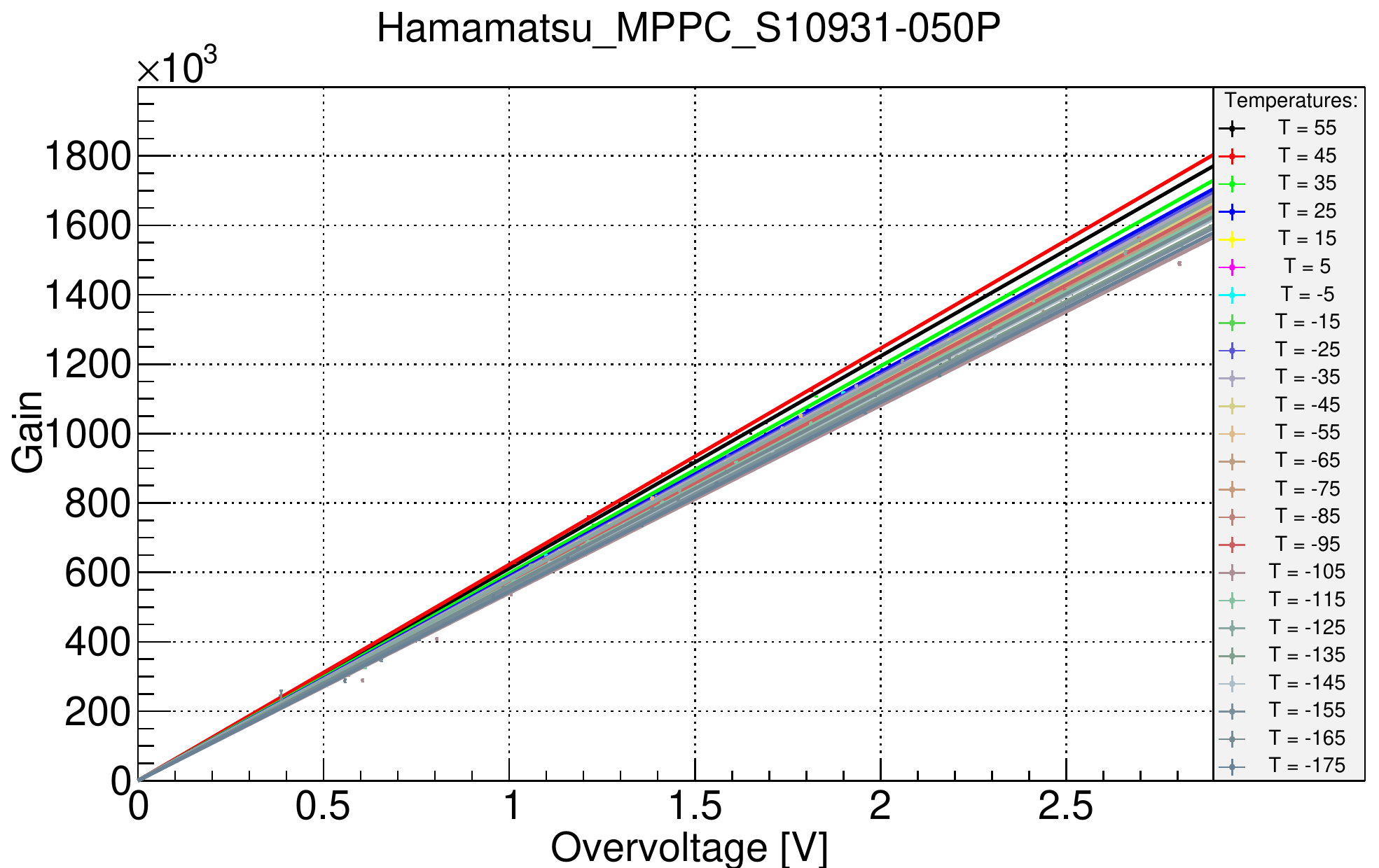}
\end{center}
\caption{Gain as a function of $\Delta V$}
\label{GainVsdV}
\end{figure}

Knowing $V_{bias}$ and $V_{BD}$ the overvoltage $\Delta V$ can be calculated as: $\Delta V = V_{bias} - V_{BD} $. 
Fig.\ref{GainVsdV}. presents the G dependence of $\Delta V$.
The gain increases linearly with $\Delta V$. Maximum G variations with T of $\sim 7\%$ are observed at a given $\Delta V$ over full temperature range used in our experiment $(225^0C)$. 
This observation is extremely useful since it shows that equivalent operational conditions, with constant G, can be obtained if the detector is working at $\Delta V$ constant independent of T. 
\section{Conclusions}
In this paper an automatic procedure for SiPM parameter analysis based on a ROOT data Analysis Framework has been described. 
In particular, this procedure allows to analyze a huge amount of experimental data and to determine SiPM parameters as gain, breakdown voltage, overvoltage, charge, signal shape (rise time and recovery time). 
As an example, the parameters of the Hamamatsu HPK detector MPPC S10931-33-050P-63813 in a temperature range from $-175^0C$ up to $+55^0C$ and different $V_{bias}$ were presented.
\section*{Acknowledgements}
The authors would like to thank Dr. Paul Rubinov and Dr. Donna Kubik for their help in the experimental setting up and data taking.

\section*{References}

\bibliography{mybibfile.bib}

\begin{thebibliography}{1}
\expandafter\ifx\csname url\endcsname\relax
  \def\url#1{\texttt{#1}}\fi
\expandafter\ifx\csname urlprefix\endcsname\relax\def\urlprefix{URL }\fi
\expandafter\ifx\csname href\endcsname\relax
  \def\href#1#2{#2} \def\path#1{#1}\fi

\bibitem{Dinu}
N.~Dinu, \href{https://tel.archives-ouvertes.fr/tel-00872318}{{Instrumentation
  on silicon detectors: from properties characterization to applications}},
  Habilitation {\`a} diriger des recherches, {Universit{\'e} Paris Sud - Paris
  XI} (Oct. 2013).
\newline\urlprefix\url{https://tel.archives-ouvertes.fr/tel-00872318}

\bibitem{probes}
F.~Daghighian, Intraoperative beta probe: A device for detecting tissue labeled
  with positron or electron emitting isotopes during surgery, Medical Physics
  21 (1999) 153.
\newblock \href {http://dx.doi.org/10.1118/1.597240}
  {\path{doi:10.1118/1.597240}}.

\bibitem{GMAPD}
R.~H. Haitz, \href{https://doi.org/10.1063/1.1713636}{Model for the electrical
  behavior of a microplasma}, Journal of Applied Physics 35~(5) (1964)
  1370--1376.
\newblock \href {http://arxiv.org/abs/https://doi.org/10.1063/1.1713636}
  {\path{arXiv:https://doi.org/10.1063/1.1713636}}, \href
  {http://dx.doi.org/10.1063/1.1713636} {\path{doi:10.1063/1.1713636}}.
\newline\urlprefix\url{https://doi.org/10.1063/1.1713636}

\bibitem{root}
Root webpage, \url{https://www.root.cern.ch}.

\bibitem{integral}
R.~L. J. D.~F. Burden, \em numerical analysis (7th ed.).

\bibitem{peaksfinding}
A.~J. Owen, \em uses of derivative spectroscopy.

\end{thebibliography}

\end{document}